# A study of state variable participation in limit-cycle of induction motor

Krishnendu chakrabarty
Kalyani Government Engineering College
Kalyani-741235, India
chakrabarty40@rediffmail.com

Urmila Kar
National Institute of Technical Teachers' Training and Research
Kolkata-7000106, India
urmilakar@rediffmail.com

**Abstract—The paper presents bifurcation behavior of a single phase induction motor. Study of bifurcation of a system gives complete picture of its dynamical behavior with the change of the system's parameters. The system is mathematically described by a set of differential equations in the state space. Induction motors are very widely used in domestic and commercial applications. Single phase capacitor run induction motors are commonly used as a prime mover for fans, pumps and compressors. This paper provides a numerical approach to understand the dynamic of an induction motor in the light of bifurcation and chaos. It is seen that the dynamic of a capacitor run single phase induction can not be ascertained by the profile of a single state variable. This paper also attempts to discuss the bifurcation behavior of the system based on the evolution of different state variables. The bifurcation diagrams drawn looking at different state variables are different in terms of periodicity and route to chaos. The knowledge of the dynamics of the system obtained from bifurcation diagrams give useful guidelines to control the operation of the induction motor depending on the need of an application for better performance.**

## I. Introduction

Single phase induction motors are widely used as a prime mover to run fan, pump, and compressor etc. for the industrial and household applications. These motors have no inherent starting torque. Single phase Induction motors have symmetrical rotor cage and non symmetrical two stator winding named as the main winding and auxiliary winding with starting or running capacitor. Both windings are supplied with same sinusoidal voltage.

Study of bifurcation and chaos has become a subject of active research in Electrical Engineering due to the presence of nonlinearity in most of the systems. The qualitative structure of dynamic of a system can change as parameters of the system are varied. This change of the parameter may cause creation or destruction of fixed points including change of stability of the fixed points. These qualitative changes in the system's dynamic with the change of parameter of the system are called bifurcation. A bifurcation diagram shows the long term behavior of equilibrium points or periodic orbits of a system as a function of a bifurcation parameter of the system.

Chaos is aperiodic long term behavior of deterministic nonlinear systems. Nonlinear systems exhibit bifurcation and chaos for certain range of parameter values.

Chaos in induction motor drive was first observed in 1989 by Kuroe and Hayashi [1]. It was pointed out that the motor torque would change in a chaotic manner when parameters of the motor were set at certain value. Bifurcation and chaos in tolerance-band PWM controlled inverter fed drive was analyzed in [2]. Akasura el al used neural network to control chaos in induction motor drive [3]. A chaotic map to generate chaotic PWM in induction motor drive was studied. The design of a chaos based PWM for phase controlled converter for electric drive was investigated [4]. Gao and Chau used a periodic speed command to induce chaotic motion in an IFOC induction motor drive [5]. Electrical chaoization was applied to a single phase shaded pole induction motor to generate chaotic motion for cooling [6]. Ye et al applied electrical chaoization to a single phase induction motor fitted to a washing machine for better performance and better washing ability [7]. There are also reported works on bifurcation and chaos in indirect field oriented control of induction machine (IFOC). The condition of generating Hopf bifurcation in IFOG induction motor drive was analyzed by choosing estimation error of rotor time constant as a parameter [8]. Nonlinear dynamics of an IFOC controlled induction motor was analyzed by Diyi Chen et al [9].

From the survey, it is seen that there are very few reported works [10] to understand the dynamical behavior of a single phase capacitor run induction motor in the light of bifurcation and chaos theory.

In this paper, we study the dynamics of single phase capacitor run induction motor (SCRIM). It is seen that the response of different state variables are different with reference to periodicity of its waveforms. This makes bifurcation diagrams of different state variables of the same system different. To ascertain the dynamics of SCRIM, the evolution all state variables are to be observed. This information is very useful to design the controllers of the system.

## II. Mathematical model of the system

There are various models of Induction motor available [11]. The d-q or two axis models for study of transient behavior of induction motors have been well tested and proven to be reliable and accurate. The single phase capacitor run induction motor with cross section and d-q model are shown in Fig. 1. The main (M) and auxiliary (A) stator windings are distributed in the stator slots so that they form d-q axes field. The equivalent circuit of one of the most popular induction motor models based on rotor reference frame, when the d-q axes rotate at rotor speed is shown in the Fig.2. &3.[12]

.

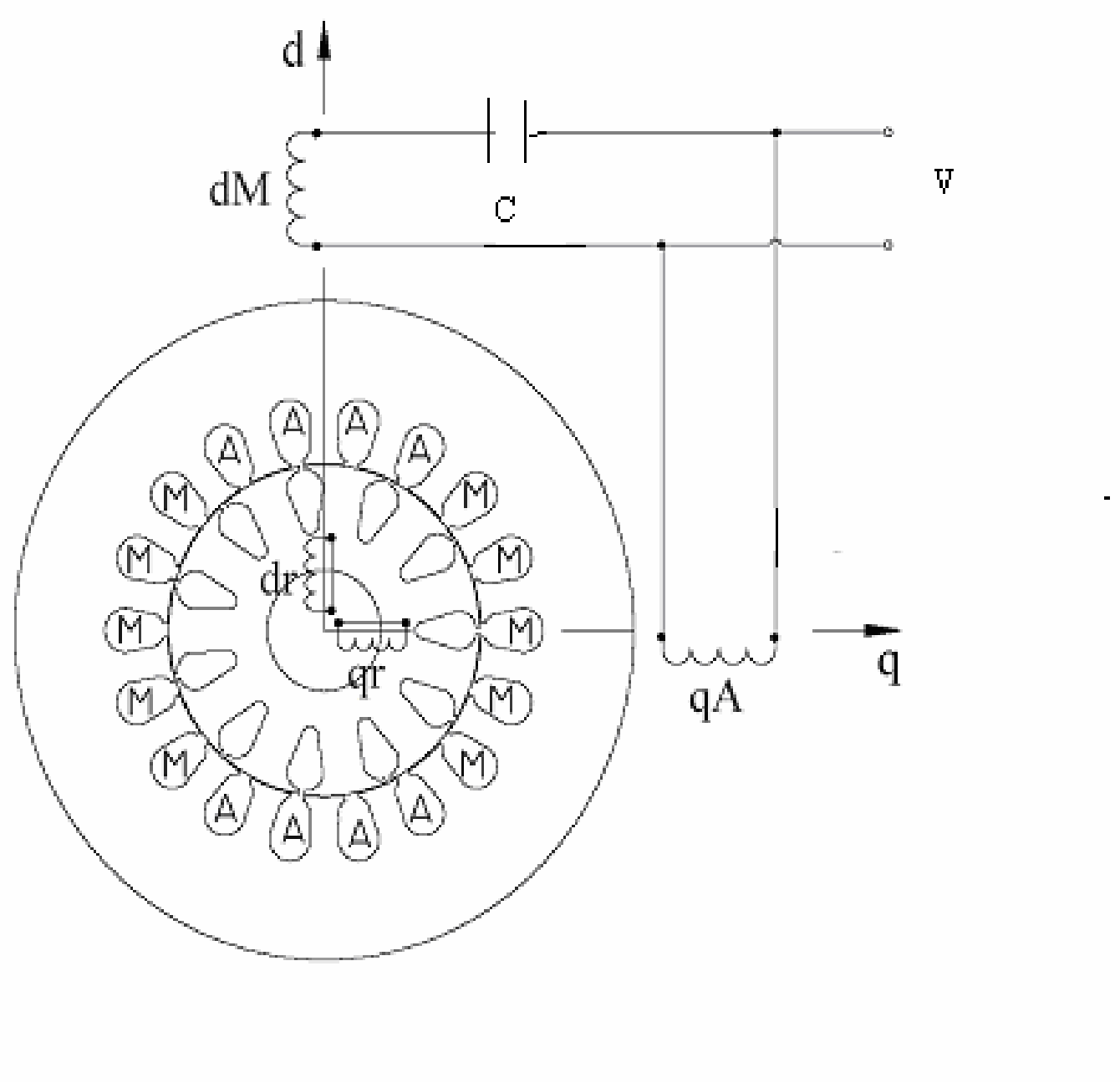

Fig. 1. View of cross section and d-q model of SCRIM

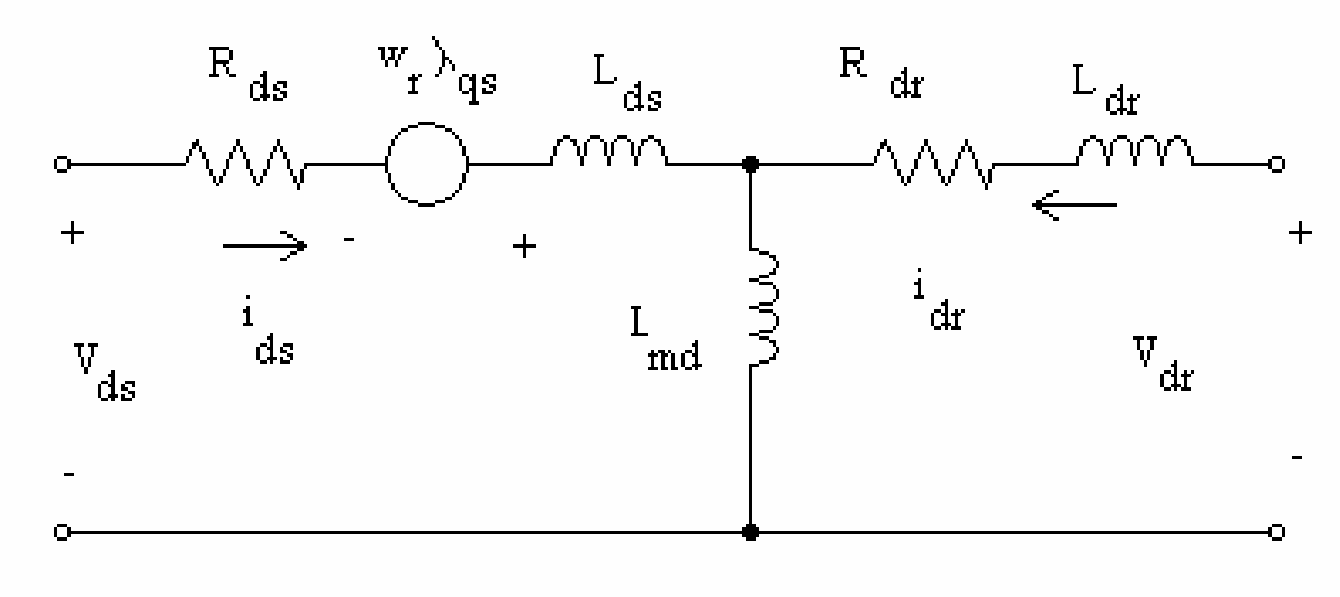

Fig. 2 Equivalent d axis model of Induction motor

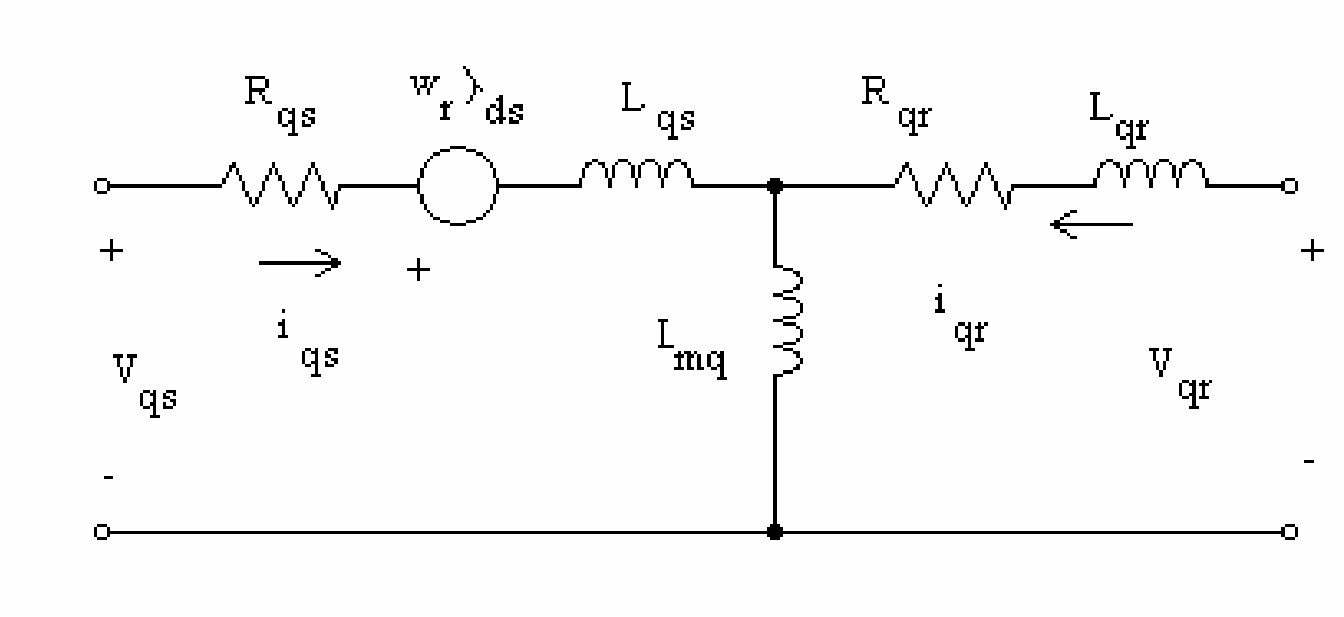

Fig. 3 Equivalent q axis model of Induction motor

The stator and rotor voltage equation are given by

$$V_{ds} = i_{ds} R_{ds} + p\lambda_{ds} - \omega_r \lambda_{qs} \quad (1)$$

$$V_{qs} = i_{qs} R_{qs} + p\lambda_{qs} + \omega_r \lambda_{ds} \quad (2)$$

$$V_{dr} = i_{dr} R_{dr} + p\lambda_{dr} \quad (3)$$

$$V_{qr} = i_{qr} R_{qr} + p\lambda_{qr} \quad (4)$$

The torque equation is given by

$$T_e = T_l + J\frac{d\omega_r}{dt} + B_m \omega_r \quad (5)$$

$$V_{qs} = V_s = V \quad (6)$$

$$V_{ds} = V_s - \frac{1}{C}\int i_{ds}\, dt \quad (7)$$

$$\lambda_{ds} = i_{ds} L_{ds} + I_{dr} L_{md} \quad (8)$$

$$\lambda_{qs} = i_{qs} L_{qs} + i_{qr} L_{mq} \quad (9)$$

$$\lambda_{dr} = i_{dr} L_{dr} + i_{ds} L_{md} \quad (10)$$

$$\lambda_{qr} = i_{qr} L_{qr} + i_{qs} L_{mq} \quad (11)$$

$$T_e = \frac{p}{2}(\lambda_{ds} i_{qs} - \lambda_{qs} i_{ds}) \quad (12)$$

$$L_{ds} = L_{lds} + L_{md}$$

$$L_{qs} = L_{lqs} + L_{mq}$$

$$L_{qr} = L_{lqr} + L_{mq}$$

$$L_{dr} = L_{ldt} + L_{md}$$

where $R_{dr}$ = Direct axis rotor resistance, $R_{qr}$ = Q-axis rotor resistance,, $L_{lds}$ = Direct axis stator leakage inductance, $L_{lqs}$ = Q-axis stator leakage inductance, $L_{md}$ =Direct axis mutual inductance, $L_{ldr}$ = Direct axis rotor leakage inductance, $L_{lqr}$ =

Q-axis rotor leakage inductance, $L_{mq}$ = Q-axis mutual inductance, $\omega_r$ = Rotor angular speed, $B_m$= viscous friction coefficient, J = moment of inertia, $T_e$= Electromagnetic torque, P=pole pairs, $\lambda_{ds}$ = direct axis stator flux linkage and $T_l$ is the load torque.

After some mathematical transformation, the mathematical model of SCRIM can be described by the following electrical and mechanical differential equations with six state variables.

$$\dot{\lambda}_{ds} = \omega_r \lambda_{qr} - i_{ds} R_{ds} + V_{ds} \quad (13)$$

$$\dot{\lambda}_{qs} = -\omega_r \lambda_{ds} - i_{qs} R_{qs} + V_{qs} \quad (14)$$

$$\dot{i}_{ds} = \frac{1}{L^2_{md} - L_{dr}L_{ds}} \begin{Bmatrix} -\lambda_{ds}R_{dr} - \omega_r \lambda_{qr} L_{dr} + i_{ds} \\ (R_{dr}L_{ds} + R_{ds}L_{dr}) - V_{ds}L_{dr} \end{Bmatrix} \quad (15)$$

$$\dot{i}_{qs} = \frac{1}{L^2_{mq} - L_{qs}L_{qr}} \begin{Bmatrix} \omega_r \lambda_{ds} L_{qr} - \lambda_{qr} R_{qr} + i_{qs} \\ (L_{qs}R_{qr} + R_{qs}L_{qr}) - L_{qr}V_{qs} \end{Bmatrix} \quad (16)$$

$$\dot{\omega}_r = \frac{T_e}{J} - \frac{T_l}{J} - \frac{B_m}{J}\omega_r \quad (17)$$

$$\dot{V}_c = \frac{1}{C} i_{ds} \quad (18)$$

## III. Dynamics of the system

To illustrate the dynamics of the SCRIM, the computer simulation is carried out. The nominal system parameters are based on the following values.

$$P = 2, R_{dr} = 20\Omega, R_{qr} = 20\Omega, L_{ds} = 21.1mH,$$
$$L_{qs} = 11.7mH, L_{md} = 235.6mH, L_{dr} = 482mH,$$
$$L_{qr} = 482mH, L_{mq} = 120mH, R_{ds} = 9.5\Omega,$$
$$R_{qs} = 4.5\Omega, J = 0.00007Kgm^2, B_m = 0.00001Nm,$$
$$C = 211.5\mu F, f = 50Hz$$

The dynamics of the motor is observed with the evolution of different states variables used to describe the system. The six state variables, direct axis stator flux linkage ($\lambda_{ds}$), quadrature axis stator flux linkage ($\lambda_{qs}$), direct axis stator current ($i_{ds}$) quadrature axis stator current ($i_{qs}$), speed ($\omega_r$) and voltage across capacitor are used to explain the dynamics of SCRIM. The behaviors of different state variables are shown in Fig.4. It is observed that the frequency of speed is different from other state variables.

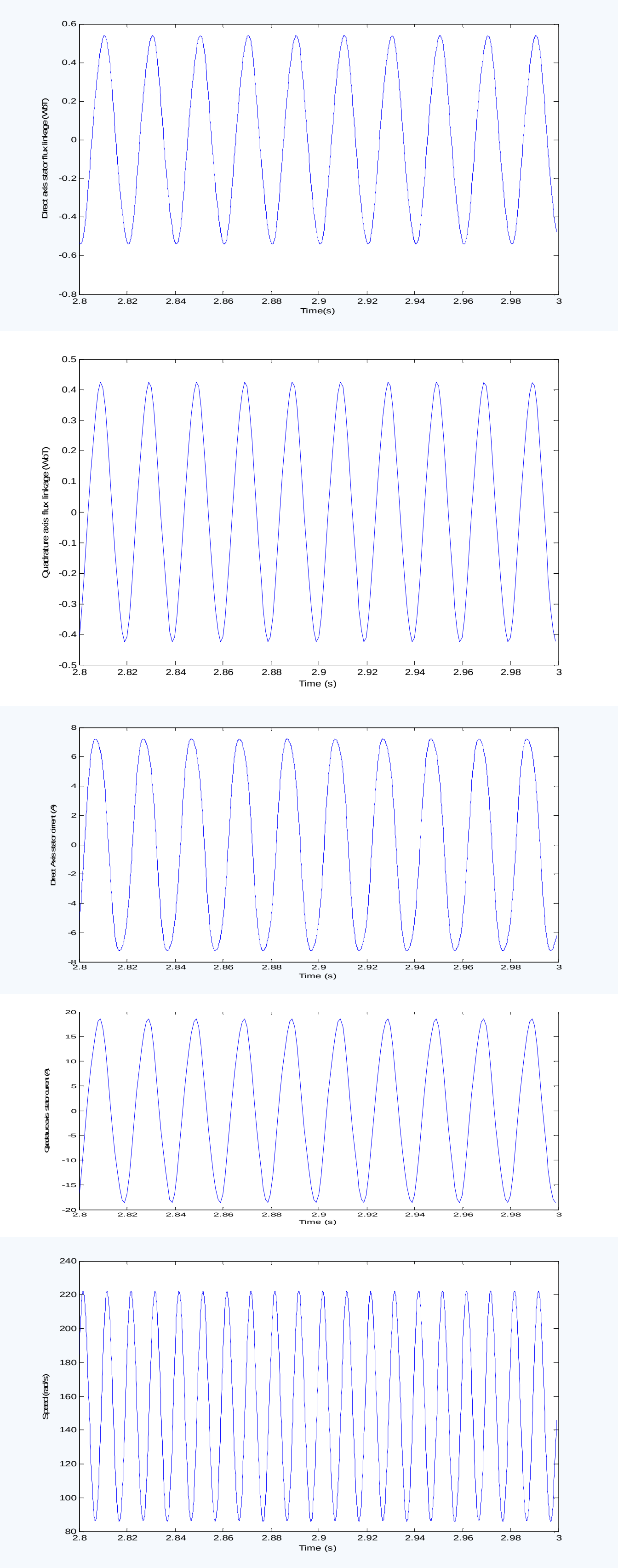

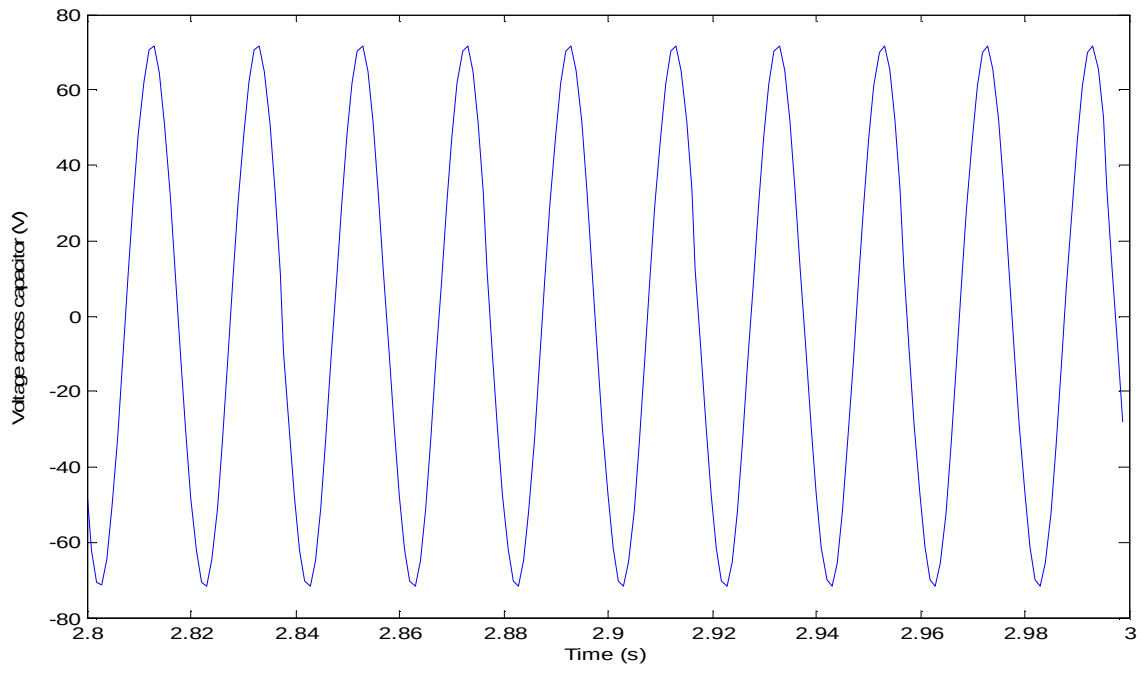

Fig. 4. Time plot of $\lambda_{ds}$ ,$\lambda_{qs}$, $i_{ds}$ , $i_{qs}$, $\omega_r$ and $V_c$ at input voltage 120 V and load torque 1.5 Nm

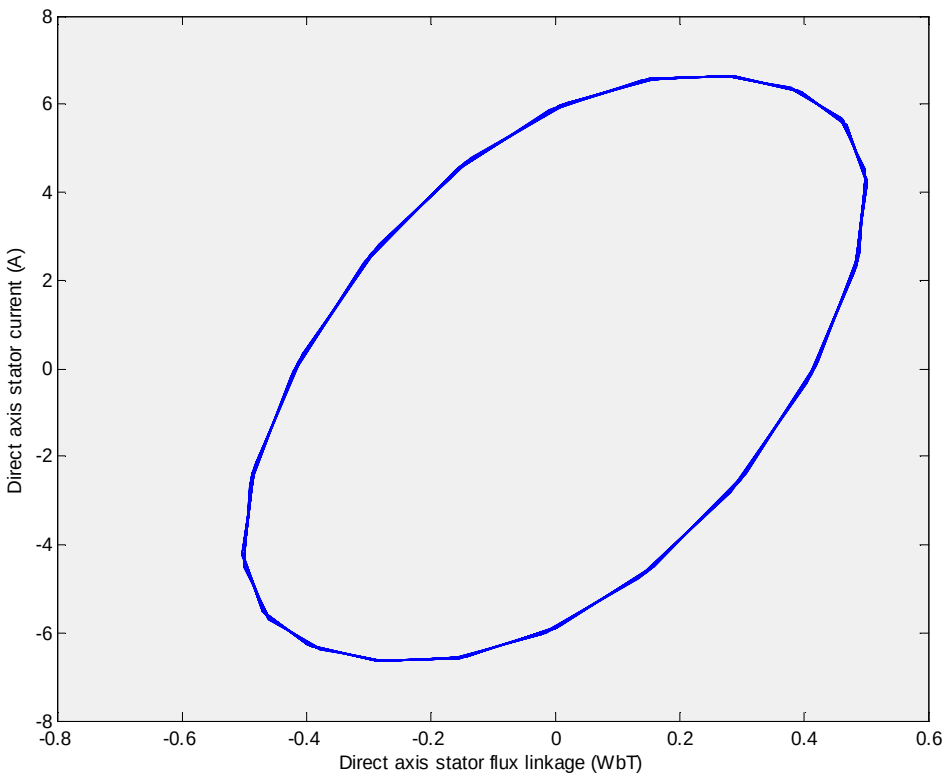

(a)

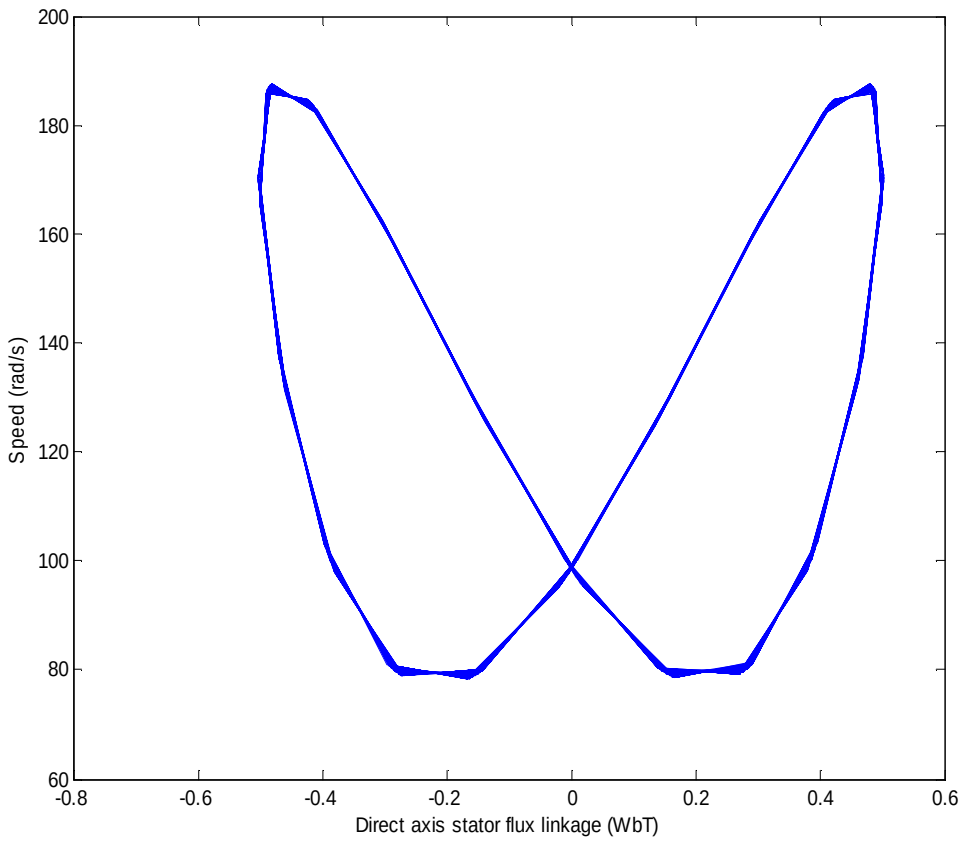

(b)

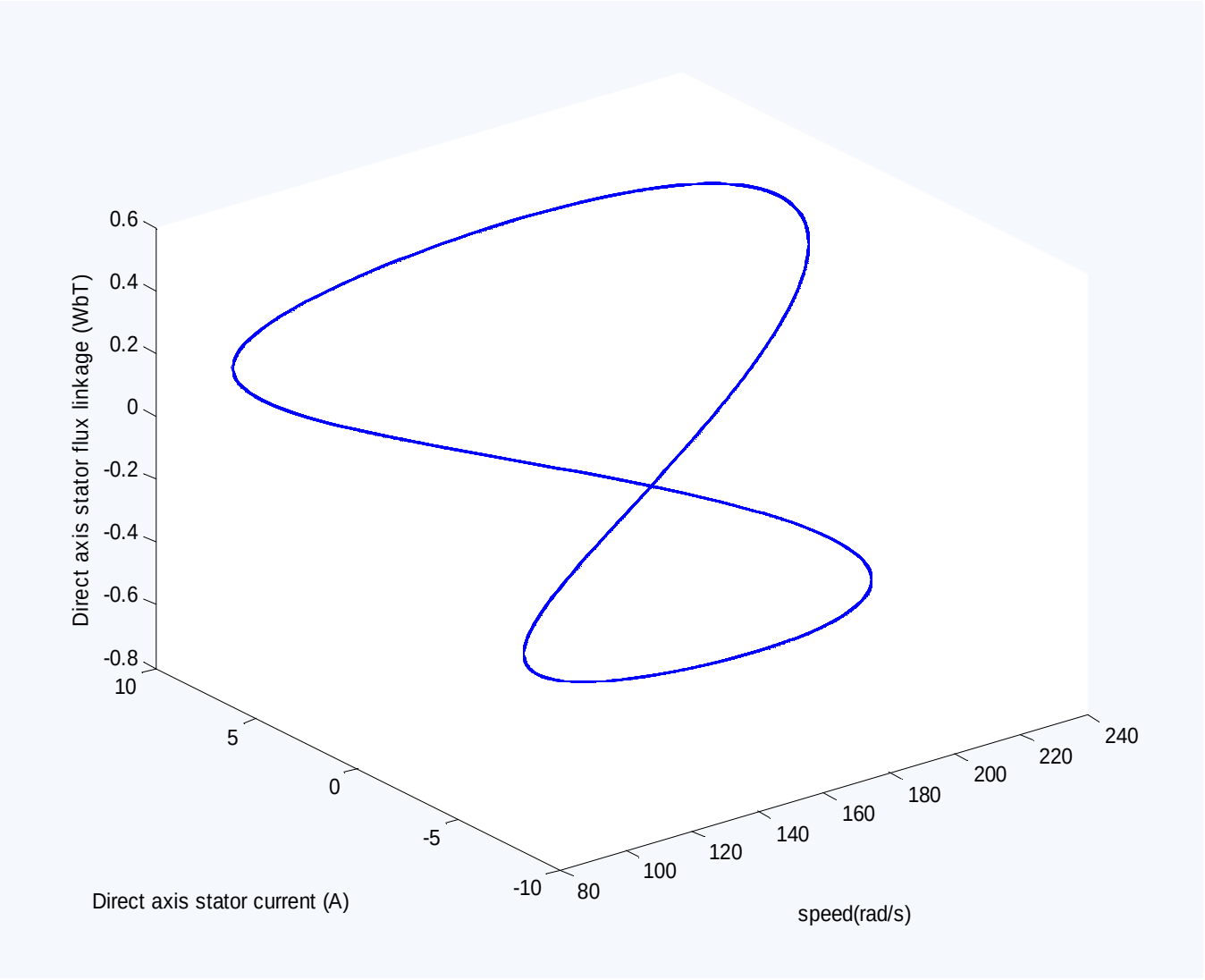

(c)

Fig. 5 (a) Phase plot of $\lambda_{ds}$ and $i_{ds}$
(b) Phase plot of $\lambda_{ds}$ and speed
(c) Phase plot of $\lambda_{ds}$ , $i_{ds}$ and speed at input voltage 120 V and load torque 1.5 Nm.

The phase plot of direct axis stator flux linkage and direct axis stator current, the direct axis stator flux linkage and speed and three dimensional phase plot of direct axis stator flux linkage, direct axis stator current and speed are shown in Fig.5.a, b & c respectively. The Fig. 5(a) shows period-1 behavior but the lower plots shows two loops in the phase plane when plotted with speed. This is due to the difference in frequency of the speed and other state variables of the system. The period of oscillation of speed and other state variables are 0.01 second and 0.02 second respectively.

Generally single loop in the phase plane determines the period-1 behavior of the system. But as the dimension of the system is more than three, and frequency of the state variables is different, it is not easy to comment on its periodicity by looking at the phase plot of the state variables. The phase plot shown in the Fig.5 (a) and Fig. 5(b) looks like a period-2 behavior.

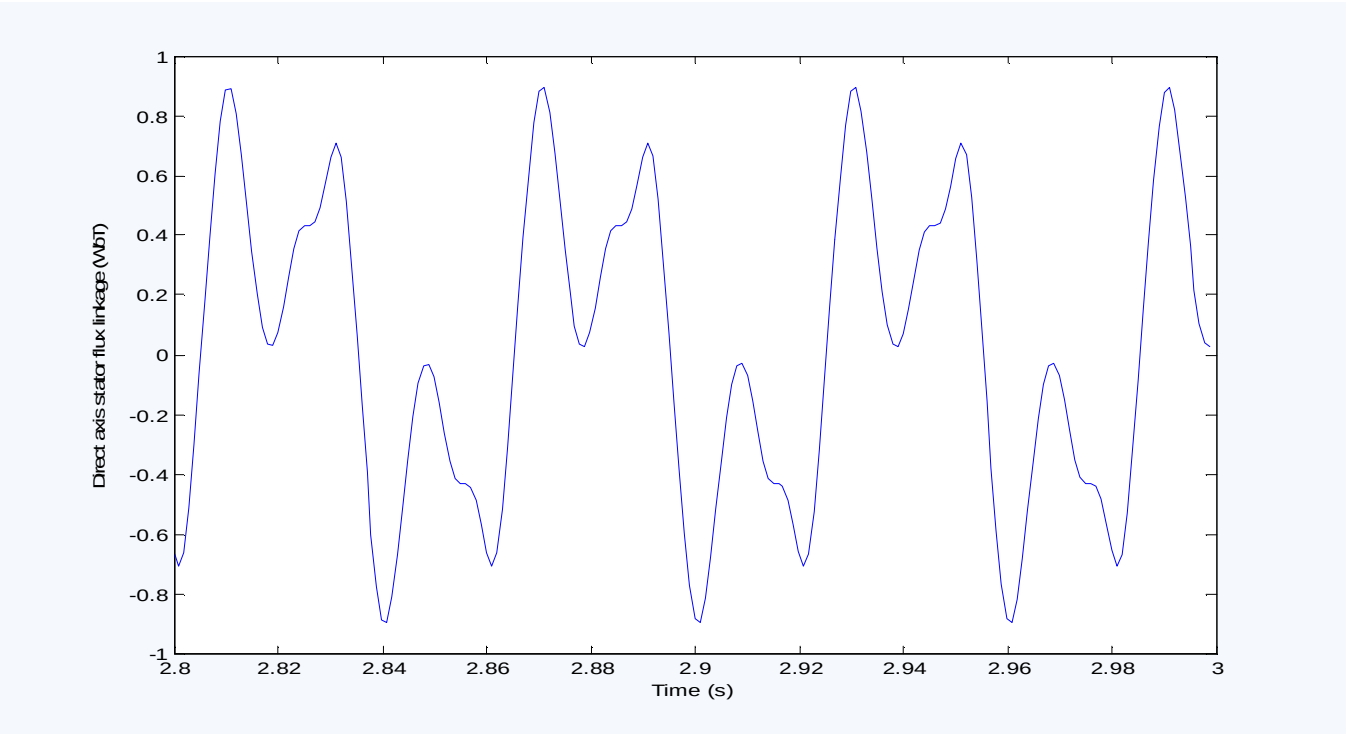

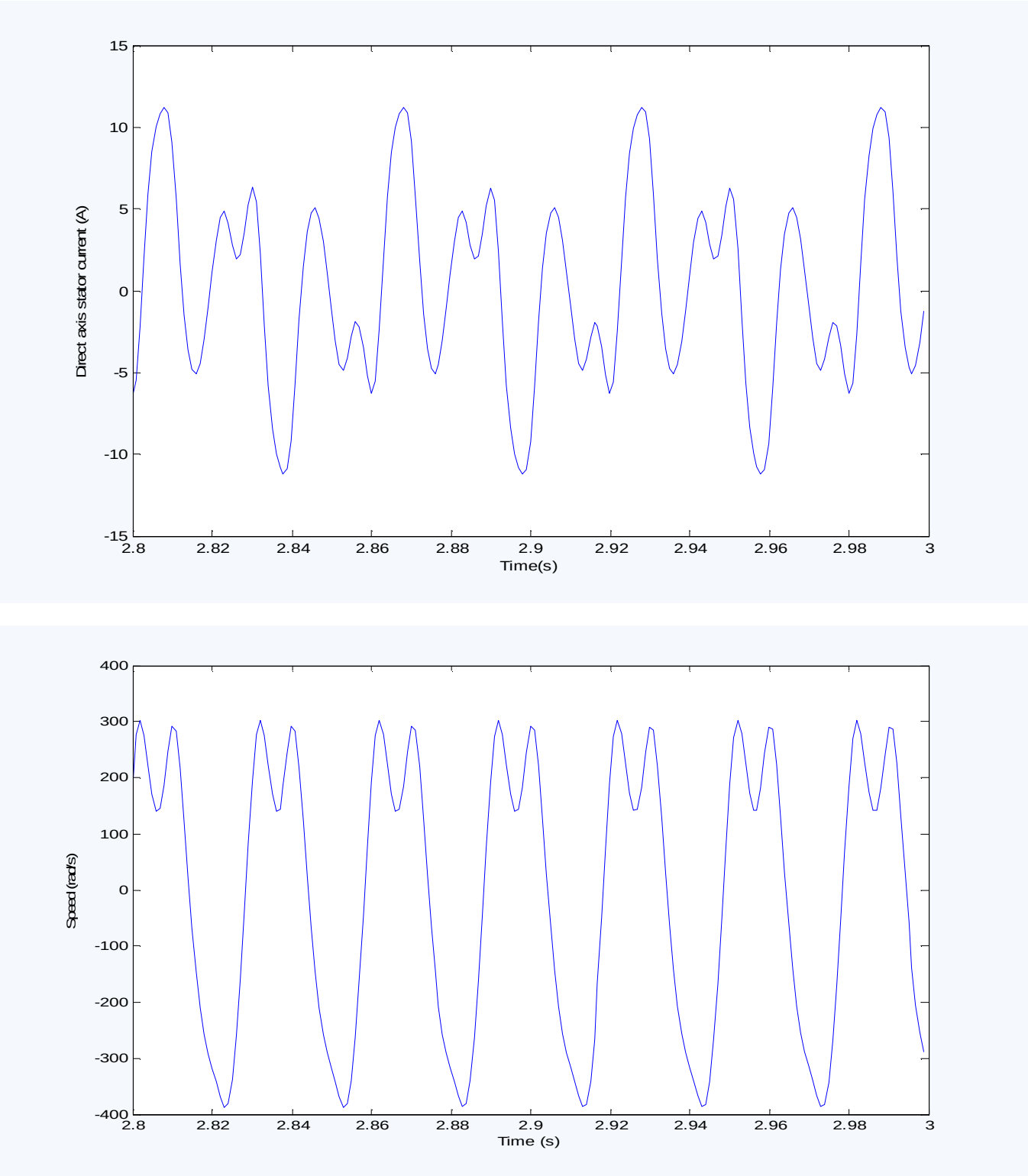

Fig. 6 Time plot of $\lambda_{ds}$ , $i_{ds}$ and speed at input voltage 160 V and load torque 1.5 Nm

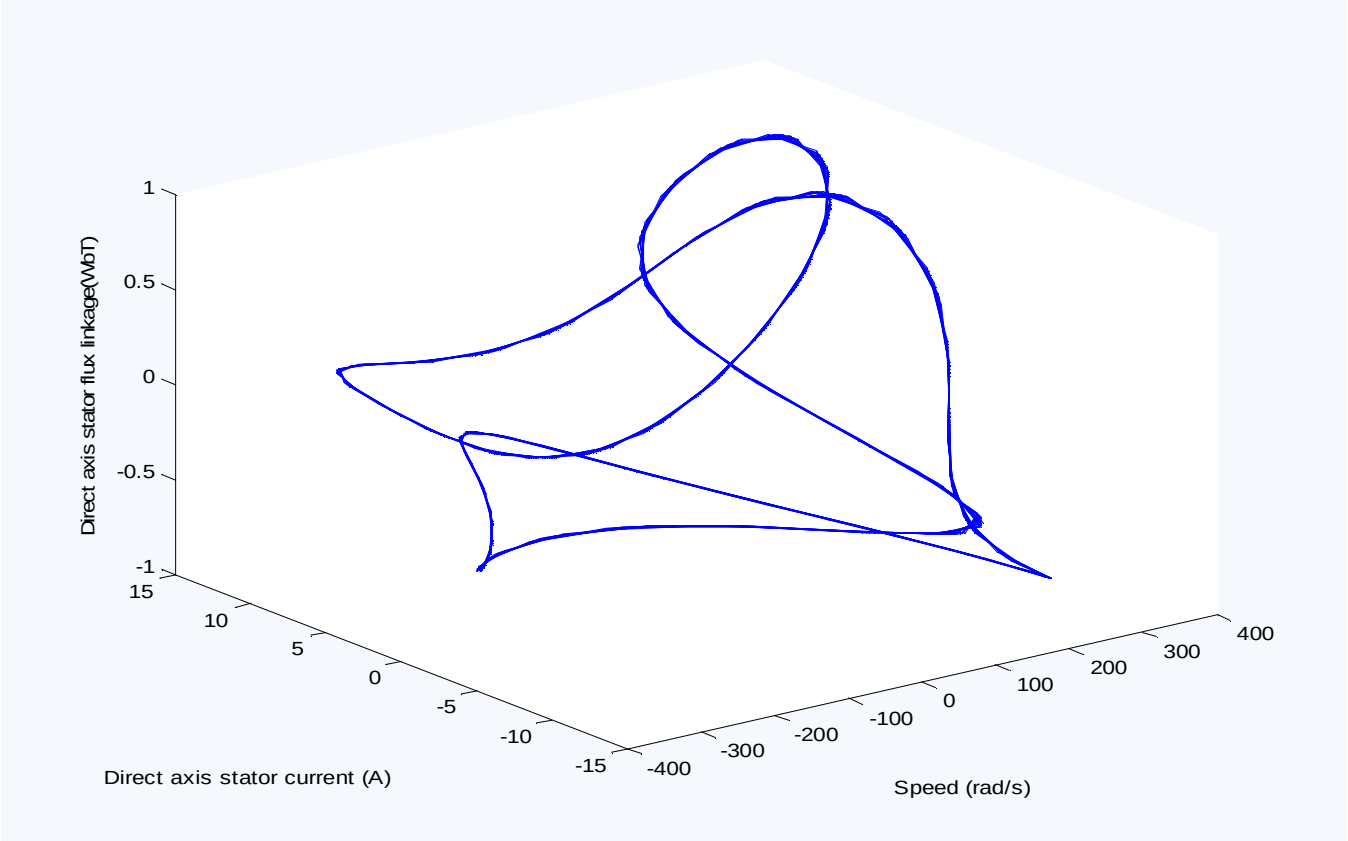

Fig.7 Phase plot of $\lambda_{ds}$ , $i_{ds}$ and speed at input voltage 160 V and load torque 1.5 Nm

But the behavior of the system's dynamic at a particular value of the parameter can not be ascertained always just by looking at the repetition of the peaks of the state variables in the time plot. It is evident from Fig. 6. If periodicity is determined by looking at the repletion of the peaks, then different state variables will indicate different periodicity. The repetition of peaks of the direct axis stator flux linkage waveform and the direct axis stator current waveform show period-5 and speed waveform looks like period-2 sub harmonic. The three dimensional phase plot of the state variables are shown in Fig. 7. But the repetition of peaks of speed and other state variables waveforms occur after 0.03 second and 0.06 second respectively at steady state condition.

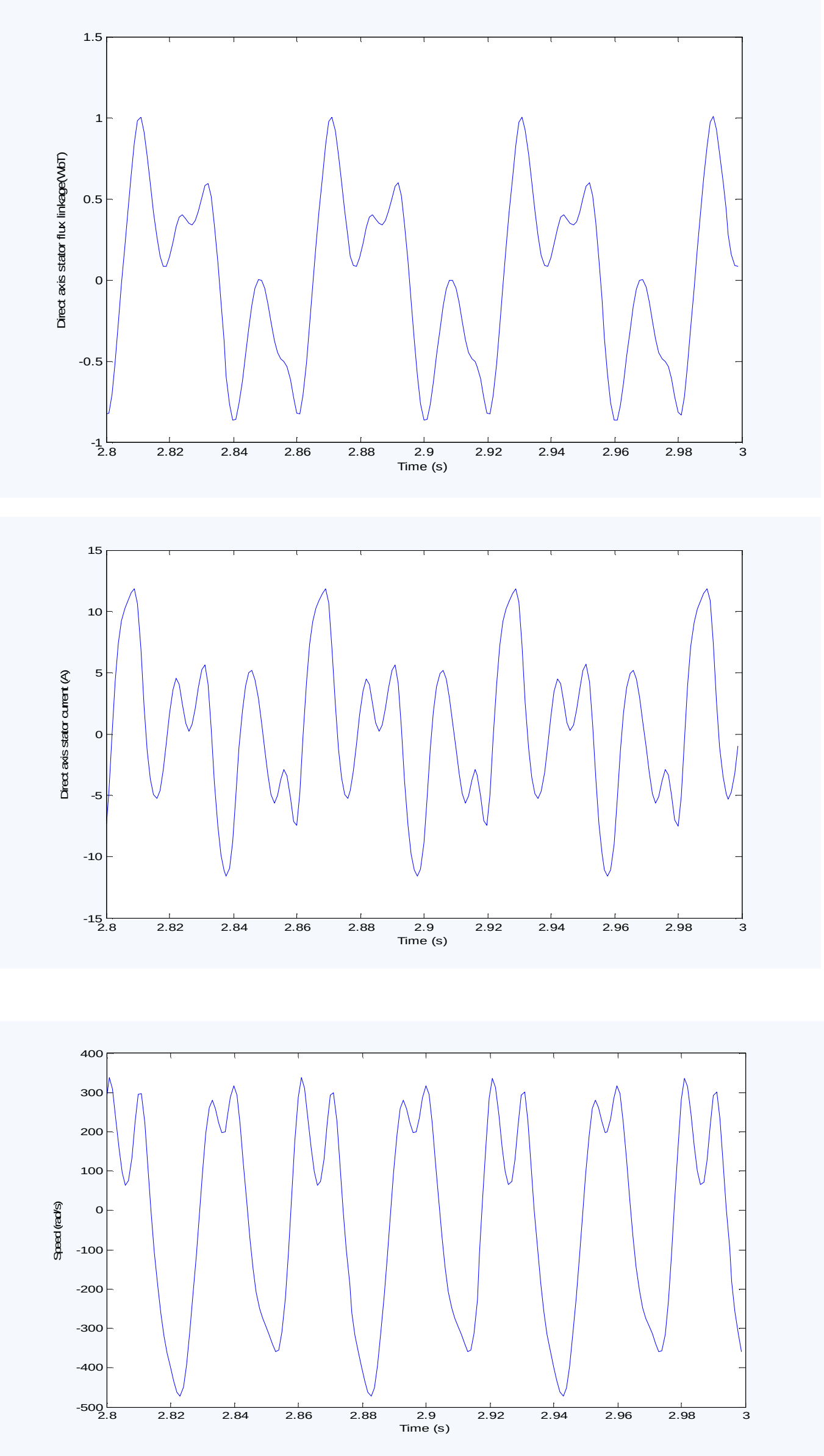

Fig. 8 Time plot of $\lambda_{ds}$ , $i_{ds}$ and speed at input voltage 165 V and load torque 1.5 Nm

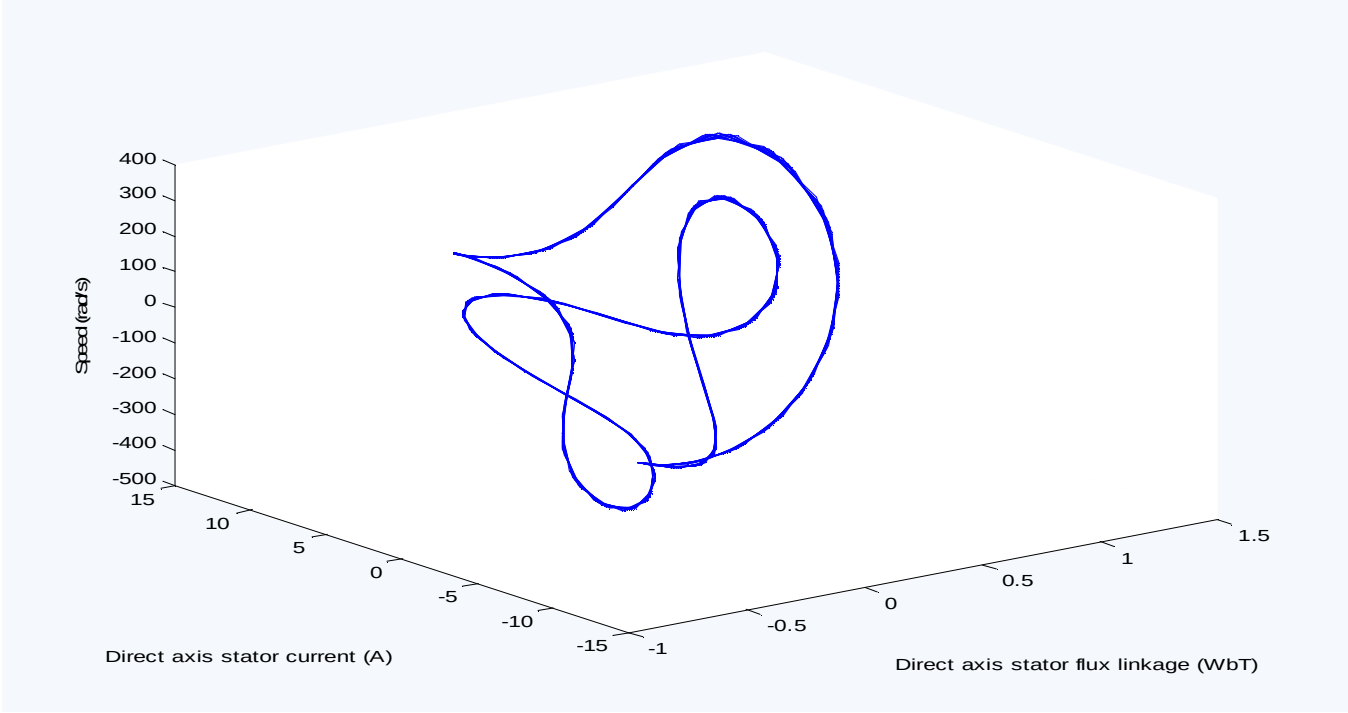

Fig. 9 Phase plot of $\lambda_{ds}$ , $i_{ds}$ and speed at input voltage

165 V and load torque 1.5 Nm

The waveforms shown in Fig. 8 are obtained at input voltage 165 V and load torque 1.5 Nm. If waveforms are sampled at the time of occurrence of peaks, one may conclude it to be period-4 looking at direct axis stator flux linkage and speed waveforms and period-5 for director axis stator current waveform. The three dimensional phase plot shown in Fig. 9. for the same can not give true information about the periodicity of the system.

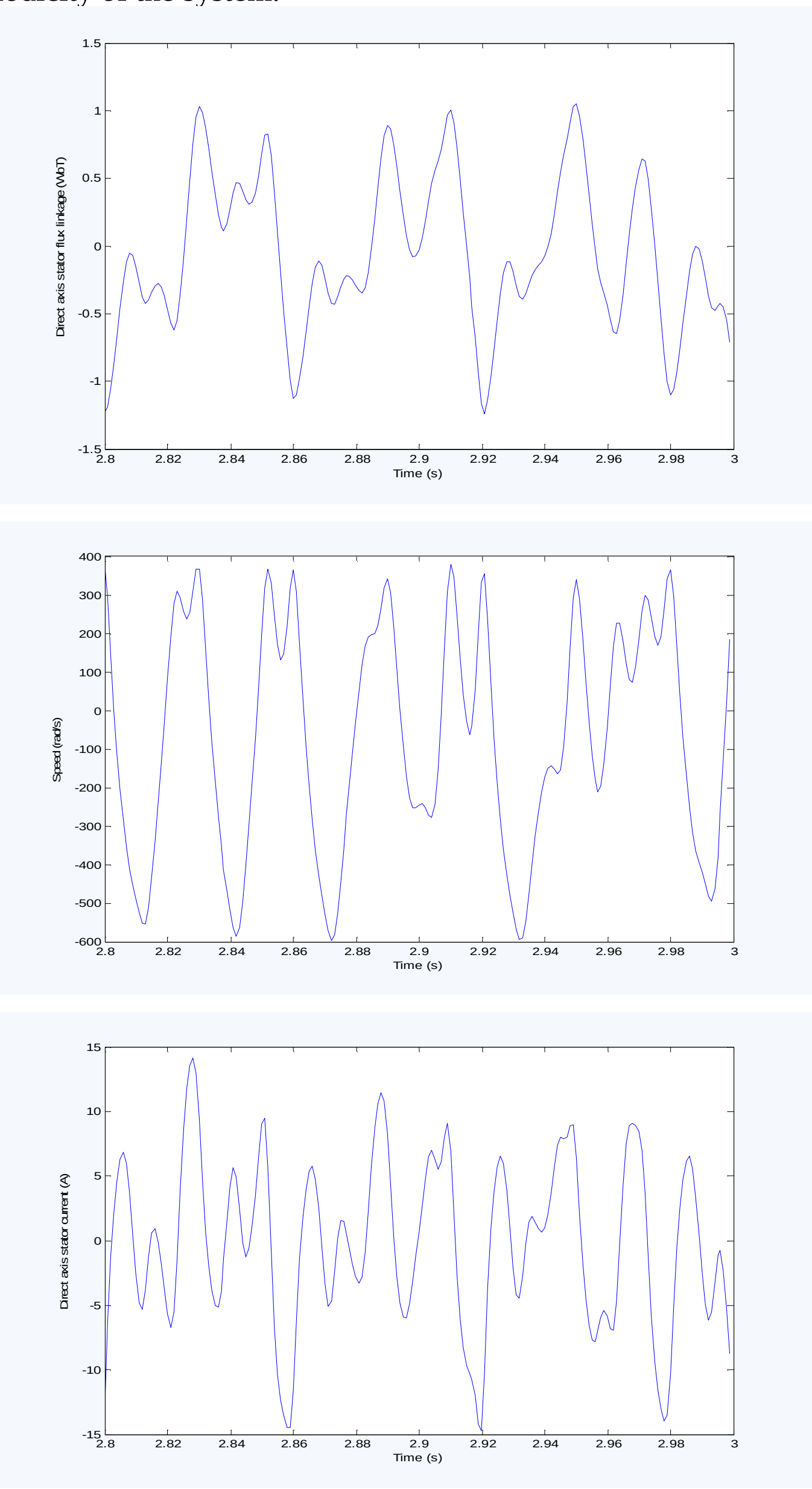

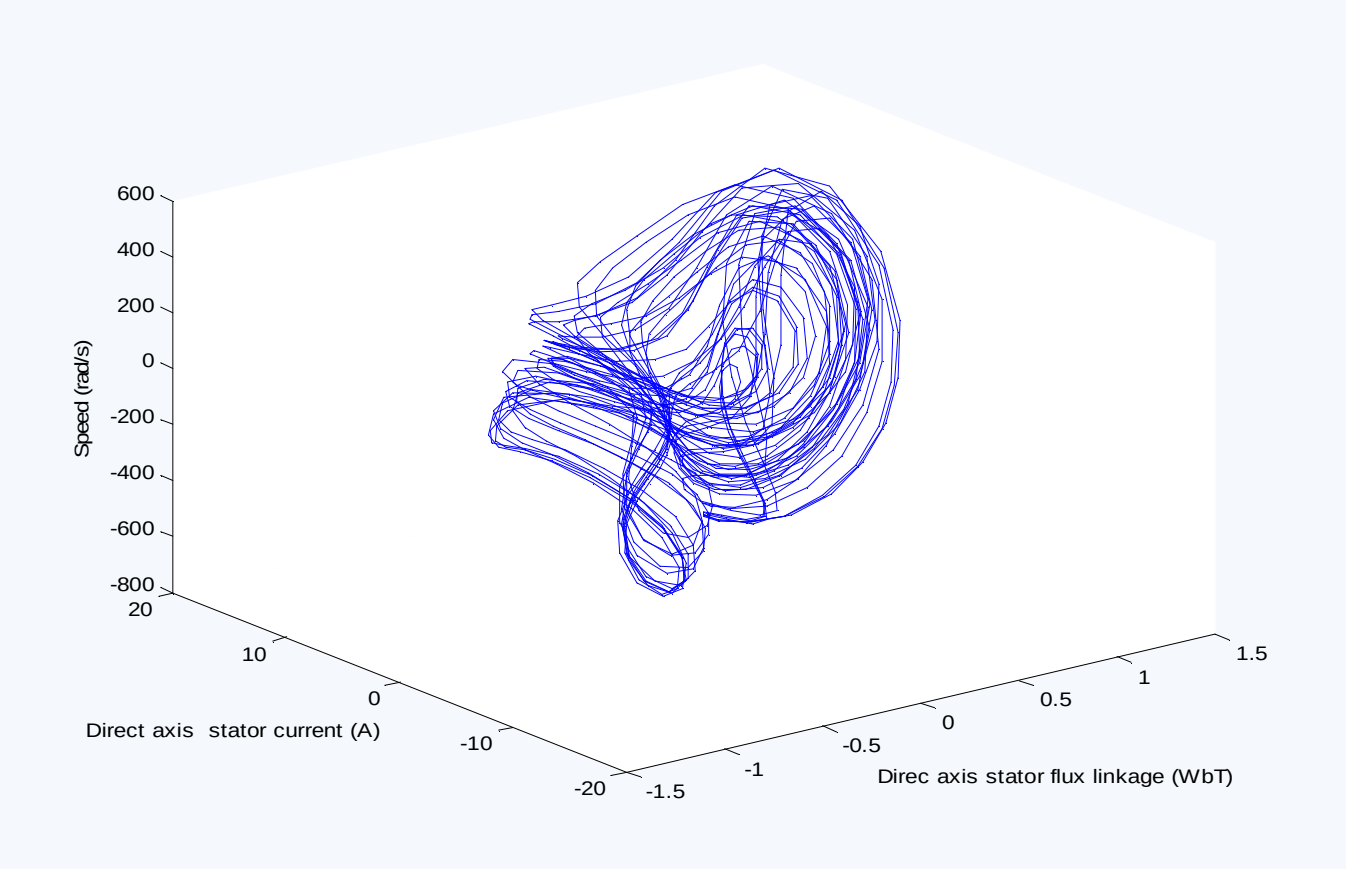

Fig. 10 Time and Phase plot of $\lambda_{ds}$ , $i_{ds}$ and speed at input voltage 200 V and load torque 1.5 Nm

Fig. 10 shows the time plot of state variables under chaotic condition at input voltage 200V and load torque 1.5 Nm.
The periodicity of the state variables can not be determined when the system goes to chaos. It is infinity under chaotic condition. For visual identification of chaotic condition, the phase plot shown in Fig. 10 would be very useful as it shows infinite loops meaning infinite periodicity.
The complete picture of the dynamic of the system in terms of periodicity i.e. sub harmonic and chaos can be obtained from the bifurcation diagram. The bifurcation diagrams are drawn observing $x_n$ of any of the state variables that describe the system at which ($t = t_n$) this state variable reaches a maximum. Bifurcation diagram with respect to four state variables, the direct axis stator flux linkage, the direct axis stator current, speed, and voltage across capacitor are shown in Fig. 11, 12 , 13, & 14.

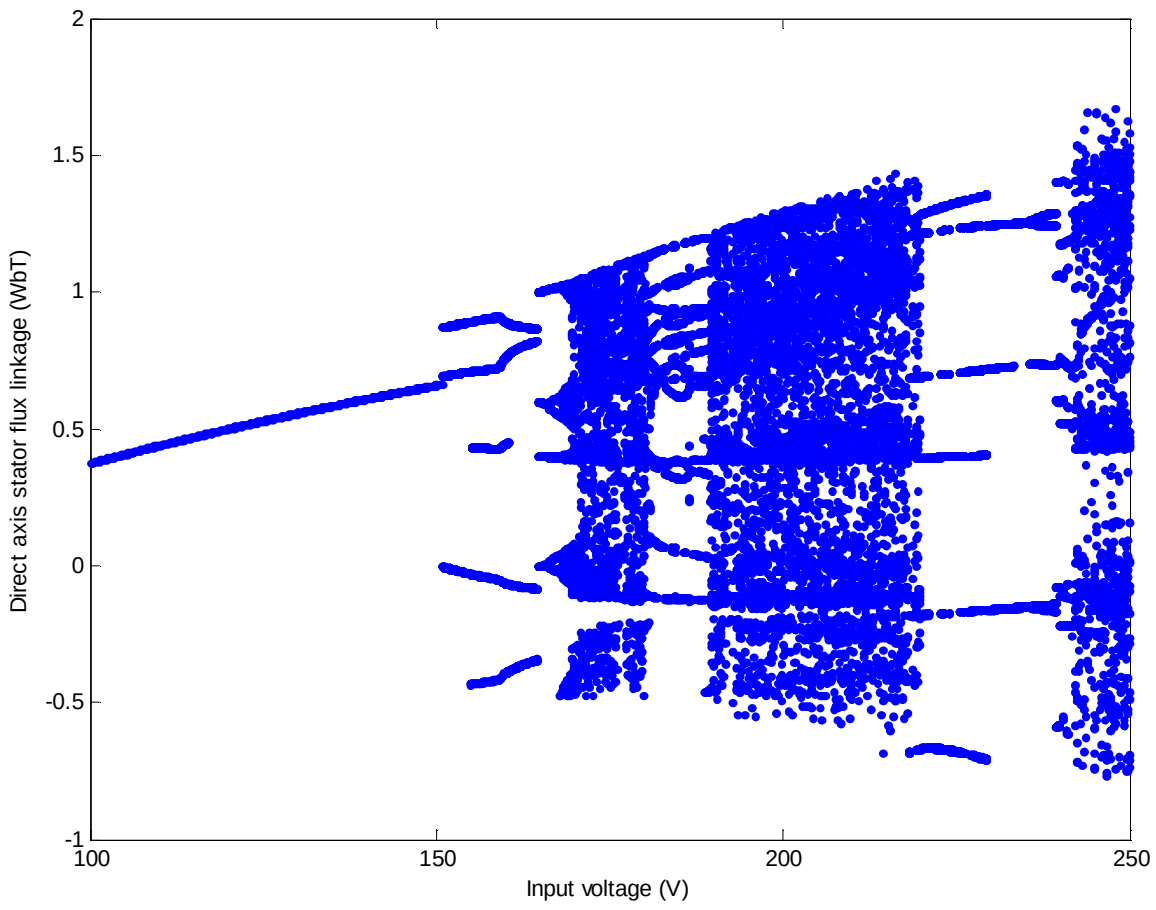

Fig.11. Bifurcation diagram of direct axis stator flux linkage for input voltage as parameter.

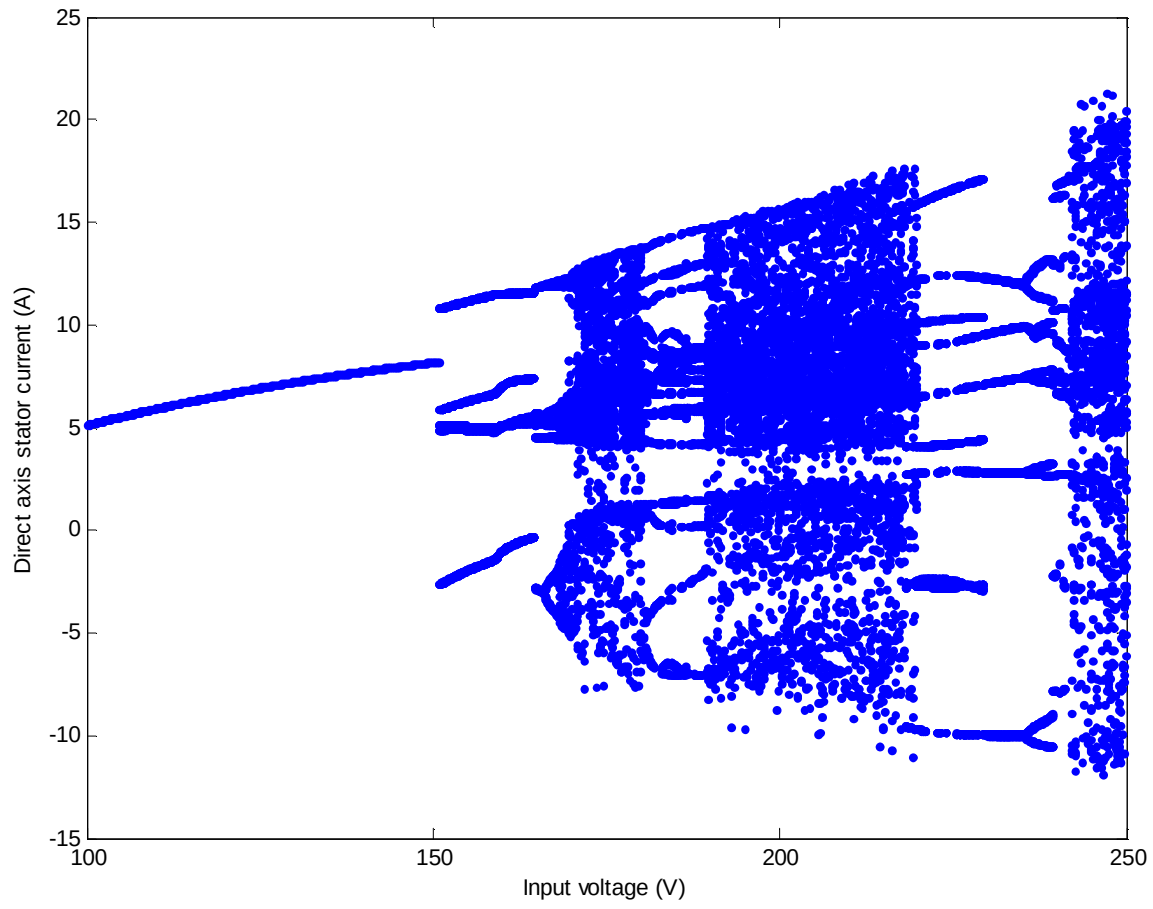

Fig. 12 Bifurcation diagram of direct axis stator current for input voltage as parameter

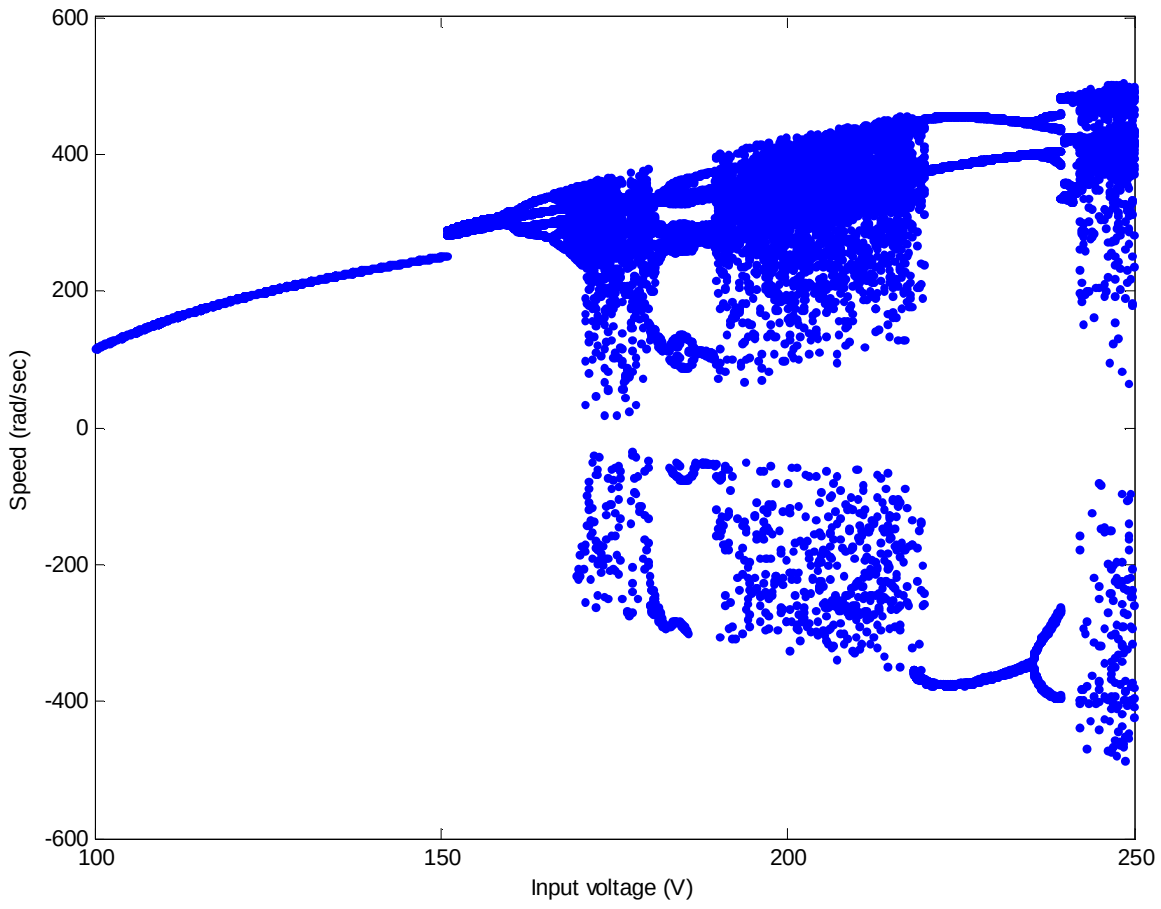

Fig. 13 Bifurcation diagram of speed for input voltage as parameter

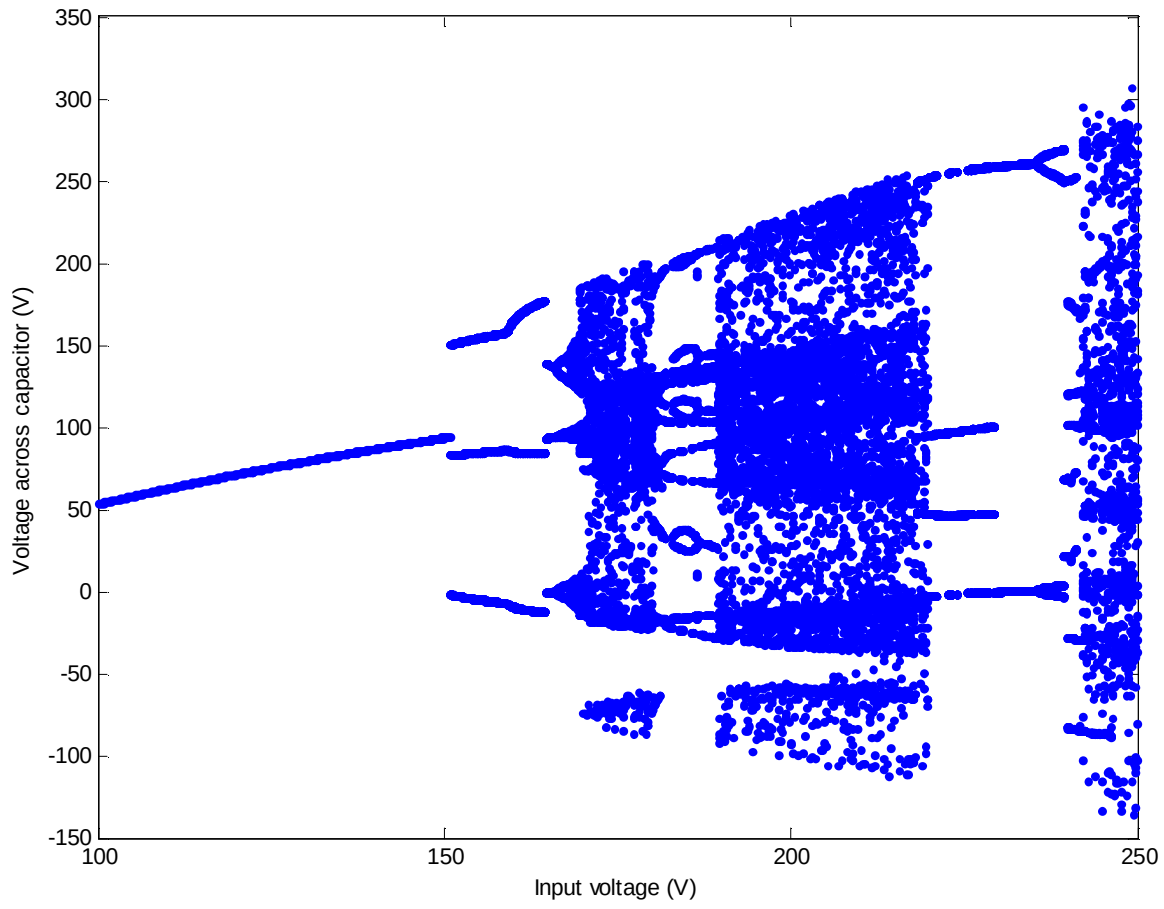

Fig. 14 Bifurcation diagram of voltage across capacitor for input voltage as parameter

It is seen from bifurcation diagrams shown above for input voltage as parameter, that all the bifurcation diagrams have same periodicity up to 150 V. Period-1 behavior is depicted when voltage is changed from 100 V to 150V. There is no smooth bifurcation from period-1, observed in all the bifurcation diagrams. Period-1 bifurcates to period-5 in Fig. 11, period-4 in Fig. 12 and period-2 in Fig. 14 and period-3 in Fig.14 respectively. Even though the chaotic zones are same for all the bifurcation diagrams, the periodic windrows in chaotic zones have different periodicity. There are existences of bubble structure in periodic windows embedded in the chaotic zone shown in Fig. 11, 13 and 14. Those structures are not available in Fig. 12.

The similar behaviors are found when bifurcation diagrams are drawn with load torque as parameters. These are shown in Fig. 15, 16, 17, 18 & 19. For low value of load torque, the direct and quadrature axis stator flux linkage show period-3 behavior, the direct axis stator current shows period-5 behavior and in the same zone, the speed shows period-2 behavior. There are similar chaotic zones present in all the bifurcation diagrams. Here also we notice periodic windows with different periodicity of different state variables embedded in the chaotic zone.

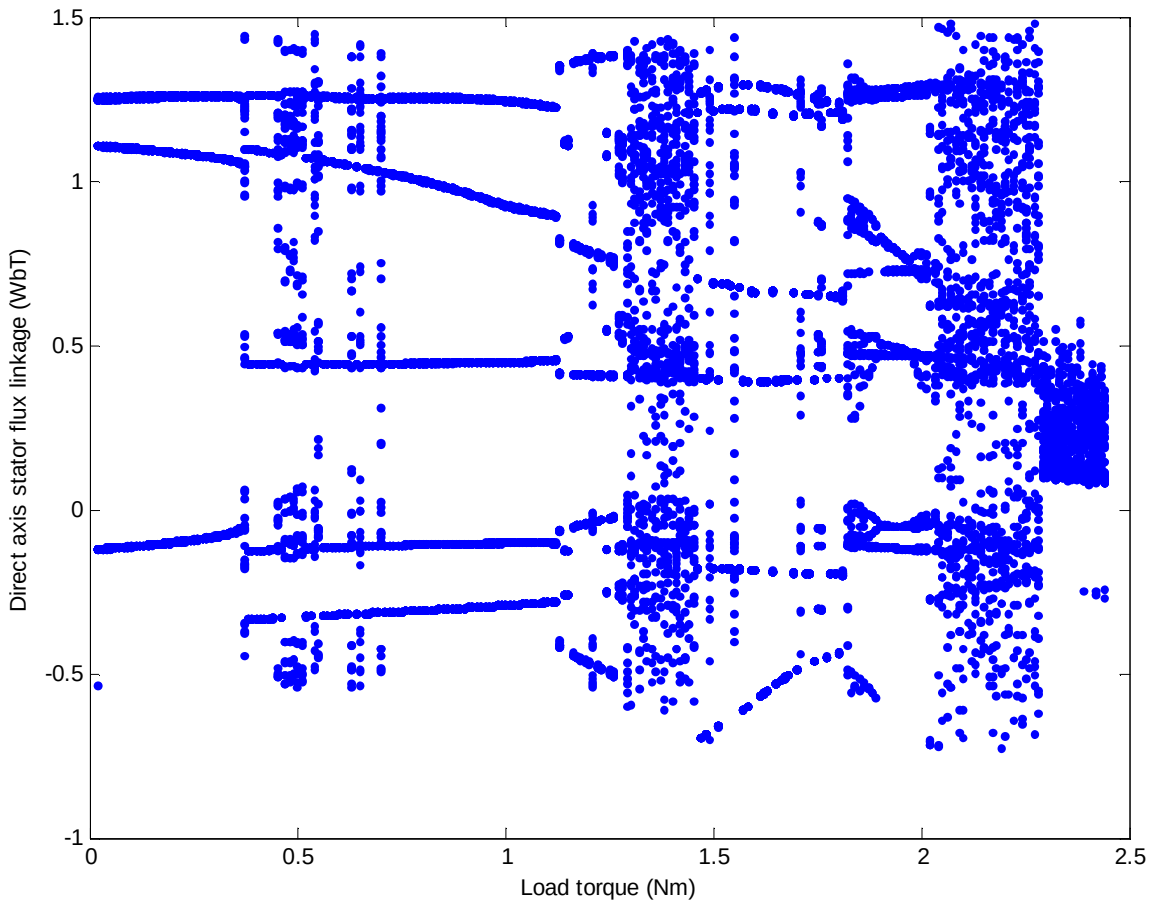

Fig. 15 Bifurcation diagram of direct axis stator flux linkage for load torque as parameter

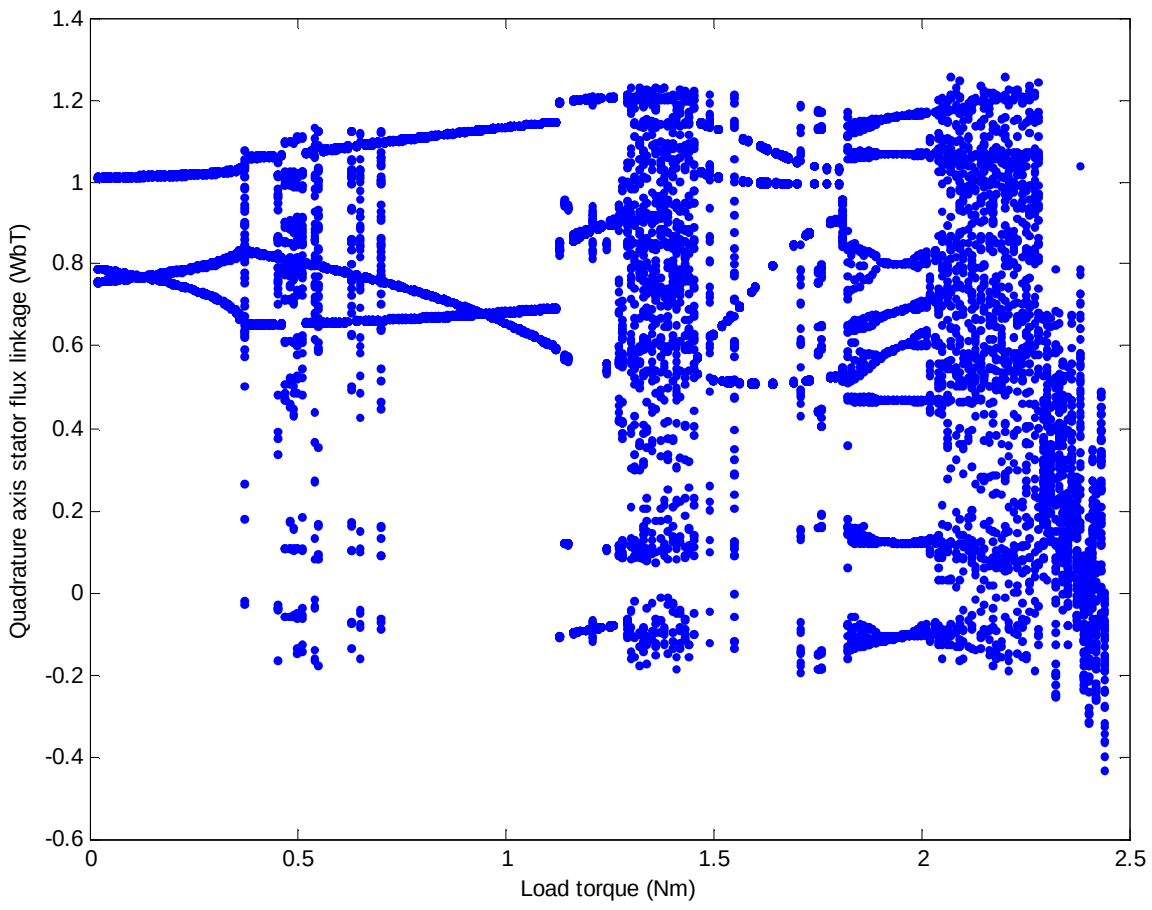

Fig. 16 Bifurcation diagram quadrature axis stator flux linkage for load torque as parameter

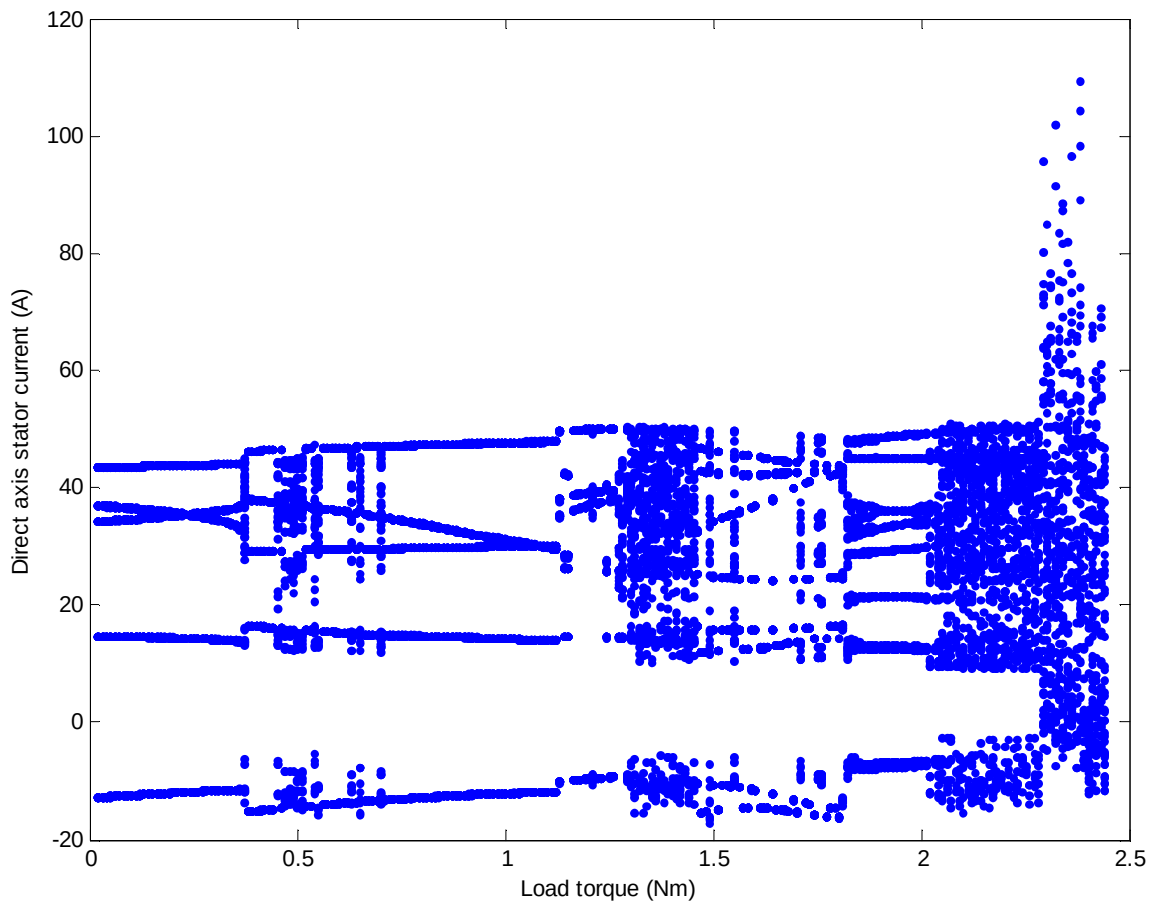

Fig. 17 Bifurcation diagram of direct axis stator current for load torque as parameter

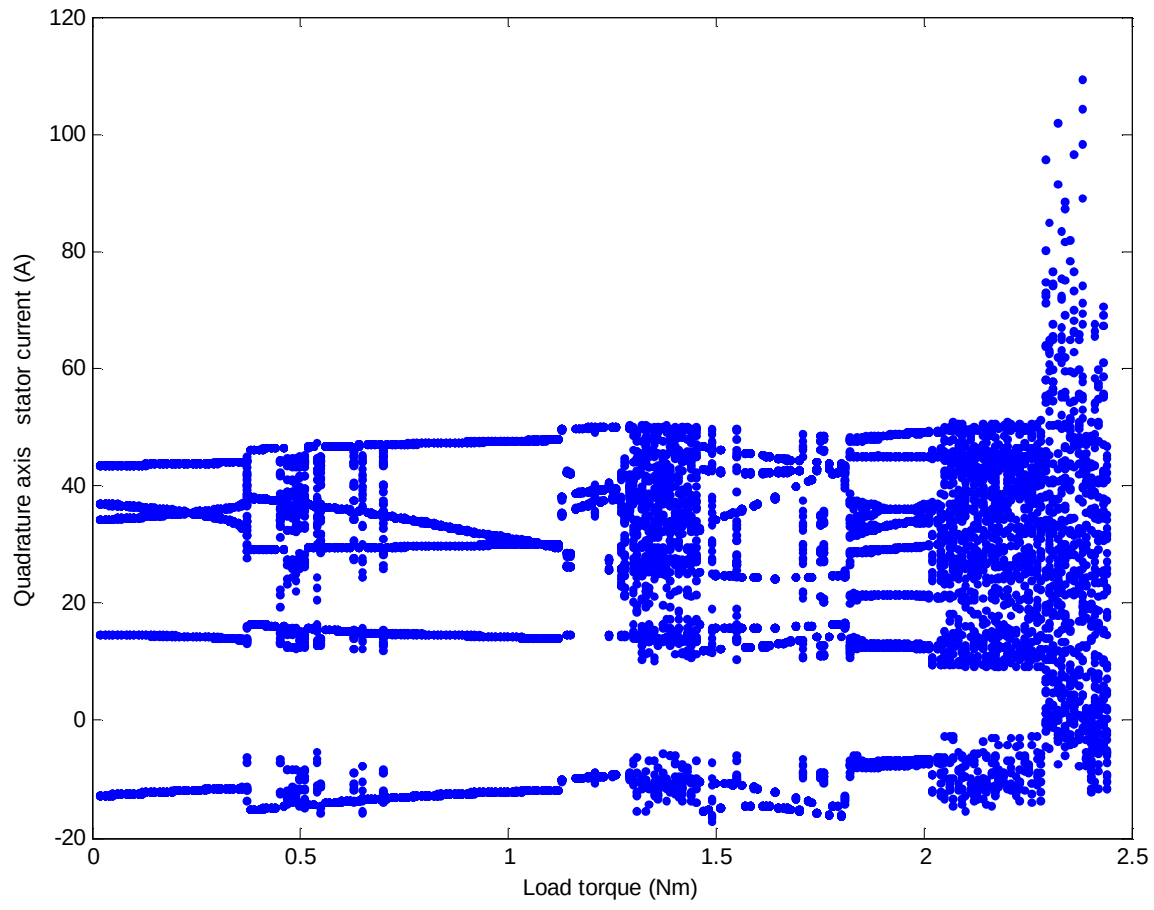

Fig. 18 Bifurcation diagram of quadrature axis stator current for load torque as parameter

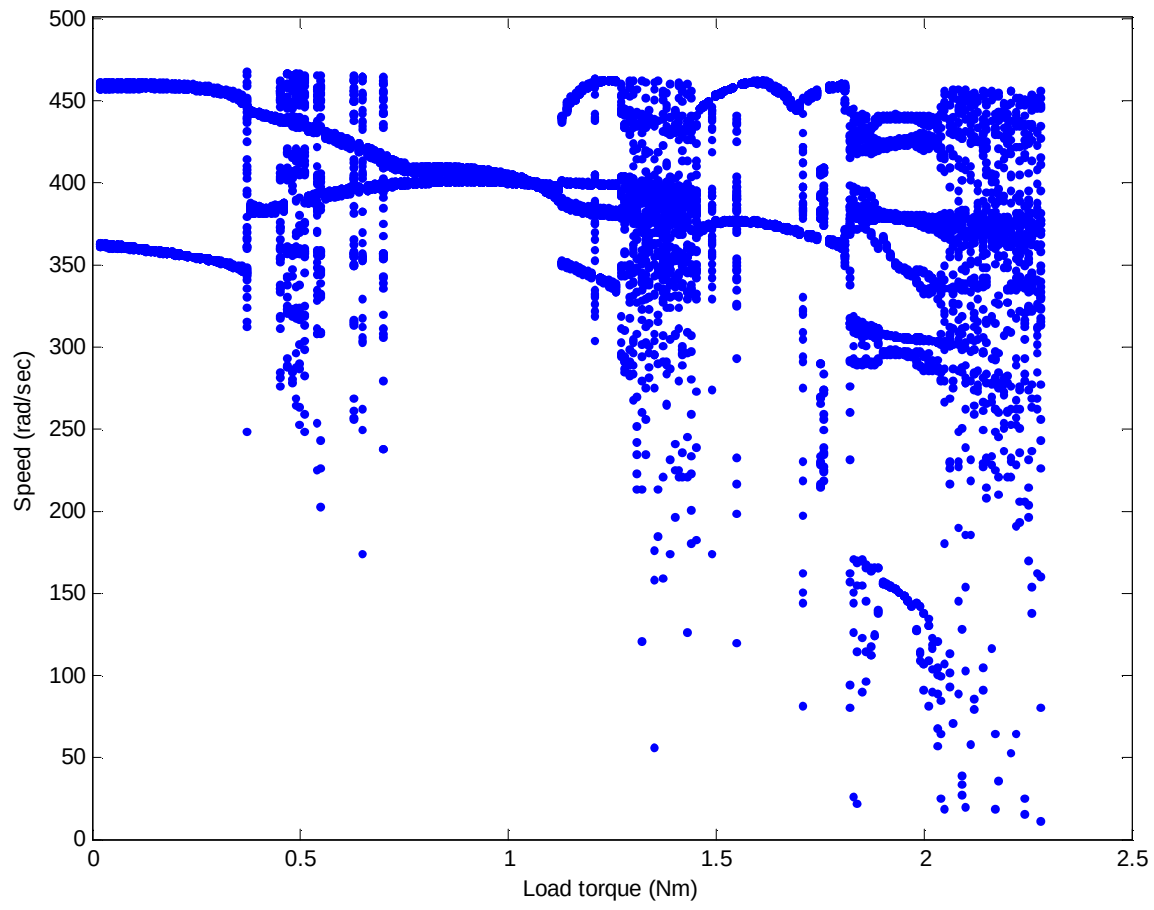

Fig. 19 Bifurcation diagram of speed for load torque as parameter.

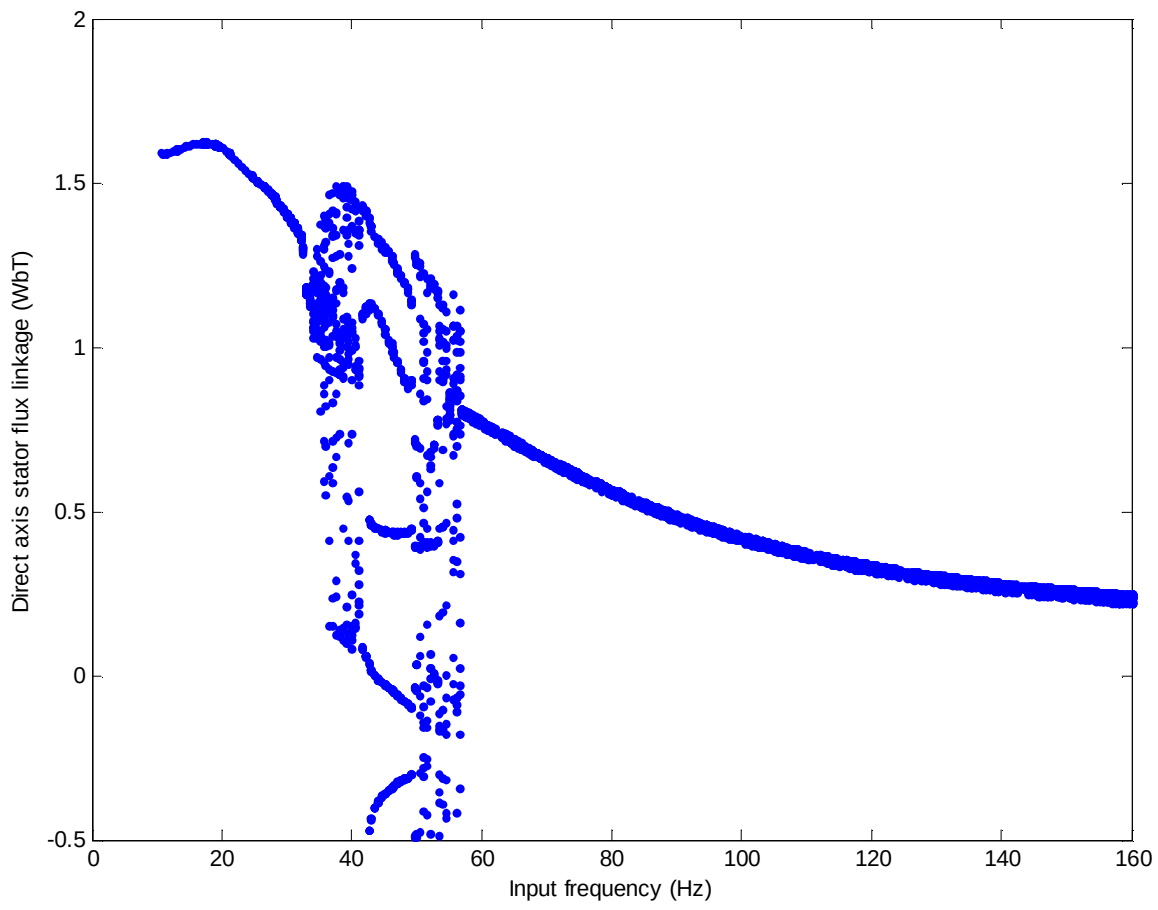

Fig. 20 Bifurcation diagram of direct axis stator flux linkage for frequency of input voltage as parameter

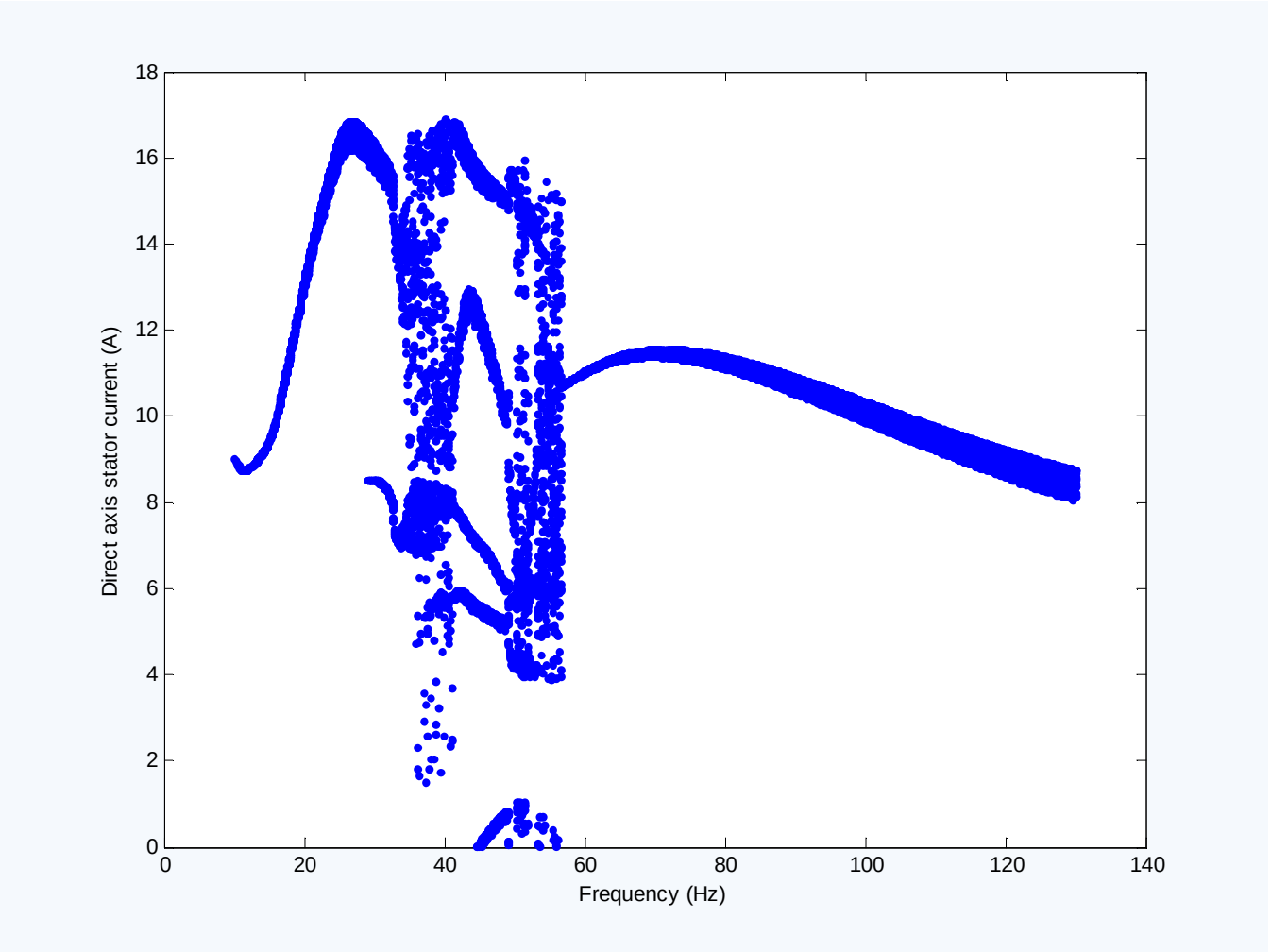

Fig. 21 Bifurcation diagram of direct axis stator current for frequency of input voltage as parameter

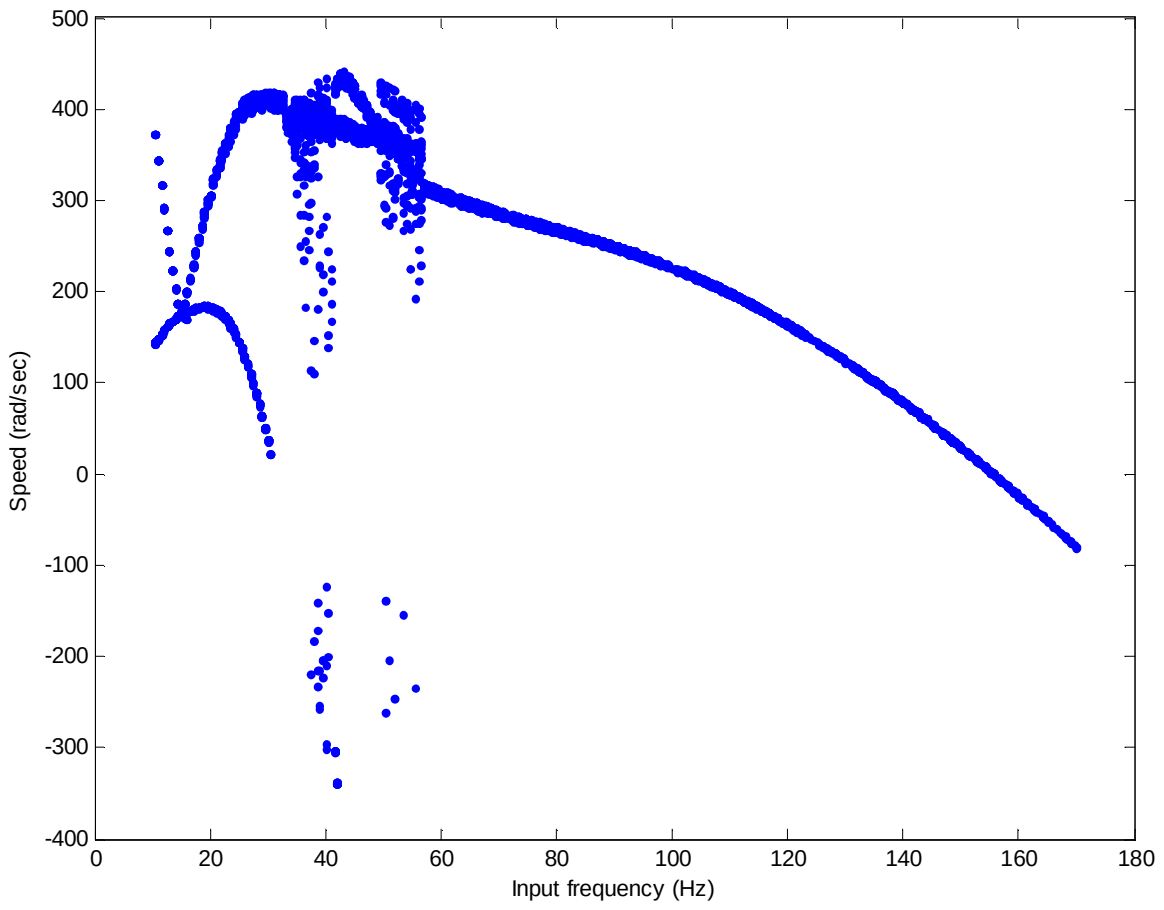

Fig. 22 Bifurcation diagram of speed for frequency of input voltage as parameter

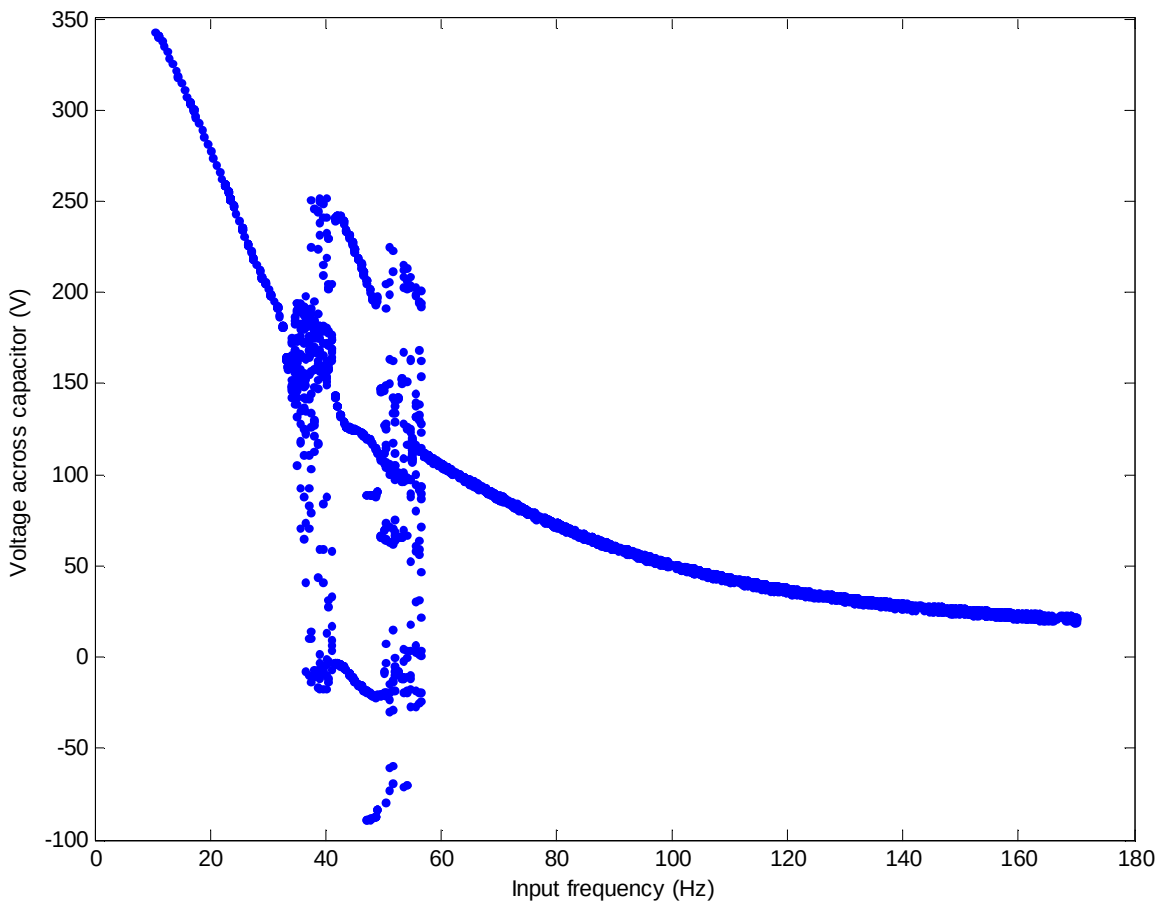

Fig. 23 Bifurcation diagram of voltage across capacitor for frequency of input voltage as parameter

When the frequency of the input voltage is taken as parameter, it is seen in the bifurcation diagram (Fig.20, 21, 22, 23) that at the lower frequency, direct axis stator flux linkage, direct axis stator current and voltage across capacitor behave as period-1 orbit. But in the same region, period-2 behavior is observed for speed. Then all the state variables converge to chaos. The periodicities of the state variables in the periodic windows of the chaotic zone are different. After 60 Hz, the entire state variables show period-1 trajectory and continues.

## IV. Conclusion

The dynamics of a single phase capacitor run induction motor has been investigated with the help of bifurcation diagrams. Generally, the bifurcation behavior is determined by monitoring the evolution of a state variable that describes the system with the change of any parameter. The general behavior of the system is ascertained from that. But in this system, it is found that bifurcation diagrams are different for different state variables. Hence it is difficult to comment on the dynamics of the system by observing the evolution of only one state variable.

Induction motors are the most used driving system, from fractional horse power to hundreds of horse power. It is shown that the dynamic behavior of the drive using induction motor as prime mover depends on many parameters. If, for some parameter values, the drive becomes chaotic, then this chaotic behavior may be desirable for some application but it may not be desirable for other applications. For normal period-1 behavior of the drive, bifurcation diagram shows the range of the parameter that would ensure the behavior. For systems like PWM controlled DC-DC converters, the range of parameter for period-1 orbit operation can be ascertained by observing the behavior of any state variable. But this is not possible in the induction motor drive. More over if the drive shows chaotic behavior for a set of parameter values, it may be necessary to achieve period-1 orbit by control of chaos. The fixed point derived from poincare map on phase plot are the basis of algorithm of control of chaos. If different state variables show different periodicity for the same parameter values, it may be difficult to find the fixed point corresponding to the unstable periodic orbit that to be stabilized for control of chaos. Moreover, it may be difficult to find the parameter space for this drive. Hence the findings of the paper will be very useful to the design engineers to choose the state variable for monitoring and control of the drive for desired output.

We acknowledge the financial support of A.I.C.T.E to complete the work.